\documentclass[12pt]{article}           
\usepackage{amsmath,amssymb}

\newcommand{\gs}{\ensuremath{g_s}} 
\newcommand{\ls}{\ensuremath{l_s}} 
\newcommand{\lP}{\ensuremath{l_{\mbox{\scriptsize P}}}} 
\newcommand{\gym}{\ensuremath{g_{\mbox{\tiny YM}}}} 

\def\p{\partial}

\newcommand{\tr}{\mathop{\rm Tr}}

\def\expec#1{\langle #1 \rangle}
\def\ket#1{| #1 \rangle}
\def\bra#1{\langle  #1 |}

\newcommand{\cN}{{\mathcal{N}}}

\begin{document}

\begin{titlepage}


\vspace*{2cm}
\begin{center} \Large \bf QCD, with Strings Attached
\end{center}

\begin{center}
Alberto G\"uijosa

\vspace{0.2cm}
Departamento de F\'{\i}sica de Altas Energ\'{\i}as, Instituto de Ciencias Nucleares, \\
Universidad Nacional Aut\'onoma de M\'exico,
\\ Apartado Postal 70-543, CDMX 04510, M\'exico\\
{\tt alberto@nucleares.unam.mx}
\end{center}

\vspace*{1cm}

\begin{center}
{\bf Abstract}
\end{center}
\noindent In the nearly twenty years that have elapsed since its discovery, the gauge-gravity correspondence has become established
as an efficient tool to explore the physics of a large class of strongly-coupled field theories. A brief overview is given here of
its formulation and a few of its applications, emphasizing attempts to emulate aspects of the strong-coupling regime of quantum chromodynamics (QCD).
To the extent possible, the presentation is self-contained, and in particular, it does not presuppose knowledge of string theory.
\vspace{0.2in}
\smallskip




\end{titlepage}

\tableofcontents

\section{Introduction}

We have long known that the richness of quantum field theory extends far beyond what can be captured with perturbative techniques, valid only when fields are nearly free. At present, we are still searching for tools that give us access to the physics of strongly-coupled fields, in as wide a context as possible. This is important for phenomenology, where high energy physicists confront strong coupling in QCD and some proposals for moving beyond the Standard Model, and condensed matter physicists face the same challenge in high-temperature superconductors and many other fascinating phases of quantum matter. It is also important purely from the perspective of formal theory, where captivating vistas have been revealed in the quest to map out the field theory landscapes, and many more surprises could still await us.

A major development in this direction was the discovery of the AdS/CFT, or gauge-gravity, or holographic correspondence \cite{malda,gkp,w}. In a limited but large class of quantum field theories, this acts as a dictionary that allows us to translate difficult questions in the regime of strong coupling into questions that are tractable, often analytically, in a different set of variables. This translation would have been useful aplenty even if we understood it only as an abstract change of variables, but becomes considerably more powerful because the new variables are physically recognizable, and therefore supply us with deep intuition for this otherwise unfamiliar regime. And what we learn seems nothing short of miraculous, because the new variables belong to a gravitational theory in a curved higher-dimensional spacetime! What is more, in the case where the field theory in question is a non-Abelian gauge theory, the gravitational description is a full-fledged string theory. The claim is that there is a complete equivalence between these two wildly different sets of variables. The gauge-gravity correspondence is a particular incarnation of the concept of \emph{duality}, which refers to any two theories that appear to be distinct but in fact turn out to be just alternative descriptions of the same physics \cite{polchinskidualities}. We say that the two theories are `dual' to each other.

This text is a relatively brief, impressionistic overview of holographic duality. It is aimed at members of the high energy community (in particular, students) with no knowledge of string theory, who want a first glimpse of the duality's origin, methods, evidence and results. There is a colossal body of work on this topic, sprawling over 12,000 papers, which we will evidently not be able to do justice to. Hopefully this brief text can serve as a pedagogic entry point to the literature. We will generally opt for citing reviews instead of long lists of original papers. There are many excellent overviews of the subject with varying degrees of detail, including \cite{magoo}-\cite{penedones}.
Good textbooks have recently become available \cite{nastase,erdmenger,mateos,zaanen}.
 We will focus on cases involving $(3+1)$-dimensional relativistic field theories with non-Abelian gauge fields, even though the correspondence is known to apply as well to other cases (e.g., \cite{giombi}-\cite{flatspace}).

When we get to applications, we will lean toward some of the ones that are motivated by QCD. The bulk of our discussion will focus on what the duality means and how it is implemented, so our presentation of specific results will be very succinct, referring to existing reviews that do a better job on this count.

It should be emphasized from the outset that holographists by no means claim to have solved QCD itself--- or any other real-world field theory, for that matter. The correspondence has clear limitations (some evolving, some rigid), and is far from an all-purpose tool. The theories we get quantitative access to are distant cousins, toy models, of the ones we would like to solve. But, as we will review in what follows, in the twenty years since its discovery, holography \emph{has} undeniably been successful in providing much needed insight into a plethora of phenomena at strong coupling. And remarkably, in various circumstances it has even led to semi-quantitative agreement with experimental data of the real-world theories that our toys are trying to model.

Gauge-gravity results complement other approaches, such as lattice QCD, which is the method of choice to carry out first-principle calculations directly in strongly-coupled QCD, but unfortunately has difficulty dealing with situations that involve time dependence or finite density. Lattice and other numerical calculations have in fact helped to verify \cite{panero}-\cite{nishimura2}
 some of the predictions that holographic duality makes for various gauge theories, starting from their alternative description involving classical gravity.
 Going in the opposite direction, there are interesting questions about quantum gravity that could be addressed via lattice studies of the corresponding gauge theories \cite{hanada}.

\section{The Basic Idea: Geometrizing the Renormalization Group}\label{basicsec}

\subsection{CFTs and RG flows}\label{rgsubsec}

Understanding a quantum field theory involves the ability to predict its behavior when probed at different energy scales. A central role in our discussion will be played by \emph{conformal field theories} (CFTs). These are theories that look exactly the same when probed at any scale. Upon first encountering this notion, it seems so removed from familiar physical phenomena that we might be inclined to believe that CFTs are as exotic as they are useless. For instance, QCD with massless quarks would be a CFT at the classical level, but quantum-mechanically, scale invariance is evidently lost due to the running of the coupling, which gives rise to the characteristic scale of the strong interaction, $\Lambda_{\mbox{\scriptsize QCD}}$. And yet, CFTs are key to our modern understanding of field theories. In fact, \emph{any} quantum field theory that is well-defined at all energies must behave like a CFT when we examine it at high energies (the UV).

 At one level this is obvious: at energies that are arbitrarily high compared to all masses or any other intrinsic energy scales that our theory might possess, the value of these scales can for all practical purposes be taken to vanish, and invariance under rescalings then follows. The subtlety is that the couplings of the theory will in general evolve as we go to higher energies, and instead of approaching constant values they might in fact diverge (possibly at a finite energy). But this option is part of what we exclude when we insist on our theory being well-defined at all energies. This is the case in particular for QCD, which by virtue of asymptotic freedom becomes in the UV a scale-invariant theory of non-interacting massless quarks and gluons. There is no association between scale-invariance and lack of interactions, so generic CFTs are not free.

 The most general definition of a quantum field theory takes a CFT as a starting point in the UV and adds to it terms that are negligible at high energies but become important at low energies. E.g., in QCD, we turn on a small value of the coupling and quark masses, and then evolve to low energies (the IR). The appropriate framework for discussing all of this is of course the renormalization group (RG), and what we have reviewed here is the fact that all UV-complete field theories can be understood as renormalization flows obtained as relevant deformations of a UV fixed point. Such deformations involve operators whose scaling dimension $\Delta$ at the fixed point satisfies $\Delta<4$. At arbitrarily low energies, one of two things happens: either the theory continues to have excitations and becomes scale-invariant again (an IR fixed point of the RG flow), or all excitations are massive, and below the lowest mass the theory becomes (scale invariant but) empty. QCD is an example of the latter possibility.

\subsection{Energy as an extra dimension}\label{energysubsec}

 Our main story in this review is closely related to these ideas: \emph{fundamentally, the holographic correspondence is a geometric implementation of the renormalization group}. The way this plays out is that upon translating to new variables, the sliding energy scale $\mu$ at which we probe the field theory, or equivalently, our spatial resolution length $\ell=1/\mu$,  becomes an  additional spatial coordinate, on a par with the original spacetime coordinates $x^{\mu}\equiv(t,\vec{x})$, which retain their meaning after the translation. The extra coordinate is often denoted $u\equiv\mu$ or $z\equiv\ell=1/u$, and referred to as the `radial' coordinate.

 We can visualize the resulting five-dimensional spacetime as analogous to an infinite stack of cards (what mathematicians and relativists would call a foliation of the spacetime), where each card (each leaf in the foliation) represents a copy of the original spacetime coordinatized by $x^{\mu}$, and height along the stack is what used to be the energy scale, but is now a new spatial direction. Information on a given card then describes the behavior of the field theory at that scale, with the UV ($u\to\infty$ or $z\to 0$) at the top of the stack, and the IR ($u\to 0$ or $z\to\infty$) at the bottom.

 Implicit in the statement that $u$ labels a spatial dimension is the idea that physics will be at least to some extent local along this radial direction. Part of this locality in $\mu$ is already familiar from the decoupling between phenomena at widely different scales in the renormalization group, and a finer degree of locality emerges under special circumstances, which will be described below. The AdS/CFT correspondence asserts that there in fact exists a complete, self-contained physical theory living on the five-dimensional spacetime, that captures exactly the same information as the original field theory: it is equivalent, \emph{dual} to it.

 \subsection{Geometrizing a CFT}\label{cftsubsec}

 Since we know that CFTs are crucial for the renormalization group, we should start by asking what a CFT becomes in the geometrized description. In addition to being invariant as usual under translations and Lorentz transformations, the vacuum state of a CFT on Minkowski spacetime R$^{3,1}$ is unchanged under rescalings (also known as dilatations) $x^{\mu}\to c x^{\mu}$. The latter map $\mu\to\mu/c$, i.e., $\ell\to c\,\ell$, so in the geometric implementation, they change the value of the radial coordinate $z$ in the same way they change $x^{\mu}$: $z\to c z$. Scale invariance of the vacuum implies that this change must leave the corresponding geometry intact. This property uniquely picks out the metric for the five-dimensional spacetime dual to the vacuum of a four-dimensional CFT. It must take the form
 \begin{equation}\label{adsmetric}
 ds^2=\frac{L^2}{z^2}\left(-dt^2+d\vec{x}^{\,2}+dz^2\right)=L^2\left(u^2(-dt^2+d\vec{x}^{\,2})+\frac{du^2}{u^2}\right)~,
 \end{equation}
which is what the relativists call \emph{anti-de Sitter (AdS) spacetime}. It has constant negative curvature, and as such, it is the simplest spacetime after Minkowski. The parameter $L$ (thus far undetermined) sets the radius of curvature.

Notice that with the metric (\ref{adsmetric}) the UV region $z\to 0$ or $u\to\infty$ is an infinite proper distance away, and in spite of this, it can be reached in finite time by signals traveling at the speed of light. So this region, known as the boundary of AdS, can affect the physics in the interior of the spacetime, referred to as `the bulk', and in defining any theory on this spacetime it will be important to specify boundary conditions at $z=0$.
Another feature of (\ref{adsmetric}) is that timelike geodesics exist that reach $z\to\infty$ in finite \emph{proper} time (i.e., our spacetime is not geodesically complete). However, this takes infinite time as measured with $t$, which is the field theory time, so one can completely describe the physics of the CFT on R$^{3,1}$ without any information of what lies beyond the surface\footnote{From the geometrized perspective, the point is that an observer at fixed $z$ and $\vec{x}$ has uniform proper acceleration, and $z\to\infty$ is an acceleration horizon for all such observers, i.e., it marks the edge of the region with which they can interact. In this respect, then, the portion of the spacetime described by (\ref{adsmetric}) (known as the Poincar\'e wedge of AdS) is analogous to the region accessible to uniformly accelerated observers in Minkowski spacetime (known as the Rindler wedge). One important difference is that uniformly accelerated observers on Minkowski feel themselves immersed in a thermal bath of temperature proportional to their acceleration, on account of the Unruh effect \cite{unruheffect}. On AdS, on the other hand, the effect of the curvature combined with the specific value of the acceleration for the observers at constant $z$ implies that the temperature associated with the horizon at $z\to\infty$ is equal to zero \cite{dl}. This is just as we expect, since (\ref{adsmetric}) is meant to describe the vacuum of our CFT.} $z\to\infty$.

Invariance of (\ref{adsmetric}) under translations, rotations, boosts and rescalings is evident; but there are four additional transformations that mix $x^{\mu}$ with $z$ and also leave the metric unchanged. The complete set of isometries (symmetries of the metric) of AdS forms the Lorentz-like group $SO(4,2)$. And indeed, CFTs in $3+1$ dimensions are invariant under exactly this same set of transformations, which for this reason is known as the conformal group. This is our first example of a match between the two descriptions. The four symmetries that we had not mentioned up to now are known as special conformal transformations, and essentially always arise in tandem with scale invariance \cite{nakayama}.

We have thus learned why AdS goes along with CFT. But so far we have considered only the vacuum $\ket{\Omega}$ of the field theory. Excited states of the same theory, of course, are not scale invariant, so they ought \emph{not} to be described simply by the empty spacetime (\ref{adsmetric}). In any CFT, there is a one-to-one correspondence between states $\ket{\psi}$ and operators $O(x)$, given by $\ket{\psi_{O}}=O(0)\ket{\Omega}$. So in the dual theory, there must be as many independent ways of placing excitations on top of the spacetime (\ref{adsmetric}) as there are independent operators $O(x)$ in the CFT. This is achieved by having a different field $\phi(x,z)$ in the new theory for each such operator, which in the state dual to $\ket{\psi_{O}}$ will be excited with a particular profile, and which admits an interpretation as a truly local field only under the special circumstances described below. Our inventory of all physical operators $O$ in the CFT includes those that are composite,\footnote{In fact, when we specialize to gauge theories only composite operators will be physical.} and similarly, in the new description the fields we have denoted $\phi$ are not necessarily elementary.\footnote{In the regime in which holography is most often studied, it is customary to refer to the $O$s as operators and the $\phi$s as fields. To avoid misunderstandings, it is perhaps worth noting that both are in fact field operators in their respective quantum theories.}
In this way we achieve a correspondence between states in the CFT and states in the five-dimensional theory. But all states in the CFT share the property that at arbitrarily high energies they resemble the vacuum, so all allowed configurations within the dual theory must reduce to empty AdS space near the UV region $z\to 0$: they must be asymptotically AdS. This is just what we would expect for states that are obtained as physical excitations of (\ref{adsmetric}), because the region near the AdS boundary has infinite volume, and it would therefore cost infinite energy to modify the behavior of the fields $\phi$ there.

\subsection{Geometrizing a generic field theory}\label{genericsubsec}

Now we can consider what a generic (UV-complete) field theory, such as QCD, should translate into. As we reviewed in Section \ref{rgsubsec}, in that case we have a renormalization flow that starts with a CFT at high energies, deformed by relevant operators, that change the physics at low energies. Even in the vacuum of such a theory, we do not have scale invariance, and therefore the dual theory, even in its lowest-energy state, must differ from (\ref{adsmetric}), incorporating some nontrivial $z$ dependence. But importantly, this dependence must turn off at $z\to 0$, because at high energies the field theory becomes conformal. The difference between this and what we described in the previous paragraph is in how fast empty AdS is approached near the boundary. Now we are changing the theory, not just the state, so we do not have the requirement that the deformation cost finite energy. We are talking then about turning on non-normalizable modes, which amounts to changing the boundary conditions at infinity. Changing the state within a given field theory corresponds instead (on both sides of the duality) to exciting normalizable modes, which are precisely the ones that can fluctuate dynamically and are quantized.

Since the spacetime directions $x^{\mu}$ are common to both descriptions, Lorentz transformation properties of the operators $O$ and their corresponding AdS fields $\phi$ must coincide. An important operator present in all local quantum field theories is the stress-energy tensor $T_{\mu\nu}$ (whose conservation follows from invariance under translations). Its counterpart in the geometrized description must be a spin two field, $g_{mn}(x,z)$, where $m,n=0,\ldots,4$, $x^m\equiv(x^{\mu},z)$.
Upon futher examination of its properties, this field turns out to be
nothing but the five-dimensional metric, or more precisely, the graviton field that describes fluctuations of the metric on top of its vacuum expectation value. This means that \emph{our original quantum field theory is dual to a theory where spacetime itself is dynamical, i.e., a theory of quantum gravity.}

Given that our direct understanding of quantum gravity is limited, up to now the correspondence has been elucidated only in cases where the gravitational description has a classical, weakly-curved limit, because only then are we able to recognize gravity as such.
Experience has taught us that two conditions are needed \cite{hpps,sundrum,fk,ahkt}. First, the emergence of a smooth, classical geometry requires that the field theory have a large number of degrees of freedom $N$ at each spacetime point, $N\to\infty$. For $N$ large but not strictly infinite, the interactions among the five-dimensional fields $\phi$, including the graviton, are controlled by $1/N$ and can thus be treated perturbatively. Even then, for the dynamics to be manageable, we need a second condition: a large separation of scales in the spectrum of the theory, so that it is a good approximation to consider only the graviton alongside a few light fields, instead of the full infinite tower of $\phi$s. In the field theory, the existence of this large gap in the spectrum (of masses or of operator dimensions) requires that the coupling approaches infinity.

The extremely fortunate fact then is that demanding some quantitative control over the gravitational description is precisely what forces us onto the strongly-coupled regime of the field theory, where we have no such control.
Conversely, the region where we do understand the field theory gives us information about the gravitational theory in a very unfamiliar regime. Use of the holographic correspondence in this direction is also very interesting, and has led to important insights into quantum gravity (e.g., \cite{elvanghorowitz}-\cite{vanraamsdonklectures}),
but lies outside the scope of this review.

The idea that the higher-dimensional description involves gravity is in consonance with two important properties of gravitational theories. First, the fact that they incorporate diffeomorphisms (diffeos) as gauge symmetries, i.e., as redundancies of the description. This is what allows equivalence with the lower-dimensional field theory, which does not possess this type of redundancy. Physical states in gravitational theories ought to be invariant under diffeomorphisms, and strictly local observables can only be discussed in the semiclassical limit. This is why above we qualified our interpretation of the $\phi(x,z)$ as local fields. Second, and related to the first point, is the fact that the entropy of a black hole is proportional to its area rather than to its volume. This can be used to argue that in a quantum theory of gravity the number of degrees of freedom inside a given region scales as one would expect for a \emph{non}-gravitational theory living on the boundary of that region. This property is known as the holographic principle \cite{thooftholo,susskind}, and with appropriate care in its formulation, it passes many checks \cite{bousso}. The generality with which this principle should apply is not yet clear, but all instances of AdS/CFT certainly constitute concrete incarnations of it. The adjective `holographic' is used because an ordinary hologram allows an essentially two-dimensional layer of film to capture in its entirety a three-dimensional image. In an analogous fashion, in the AdS/CFT correspondence we find a lower-dimensional field theory which manages to codify faithfully the physics of a higher-dimensional, gravitational theory.

\subsection{Non-Abelian gauge theories are stringy}\label{nonabeliansubsec}

As stated in the Introduction, we will focus here on examples of the correspondence where the field theory is a four-dimensional non-Abelian gauge theory. The gauge field $A_{\mu}(x)$ in these theories is an $N_c\times N_c$ matrix, where $N_c$ is the rank of the gauge group---e.g., $N_c=3$ for the $SU(3)$ of QCD. Because of our special interest in this particular theory, in all examples we will choose to refer to the matrix indices as `color' indices, which explains the $c$ subindex in $N_c$. Importantly, just as the redundancy associated with diffeos of the gravitational theory is not transcribed to the field theory, the redundancy associated with the internal local symmetry of the gauge theory is not transcribed to the gravitational description. The duality translates only physical, gauge-invariant quantities. The list of physical operators $O(x)$ to which we referred previously does not include then the elementary gluon or quark fields that we are used to dealing with in perturbation theory, but only gauge-invariant combinations of them, such as the energy-momentum tensor
or the quark and gluon condensates.
These are the objects that have counterparts in the gravitational theory. For non-Abelian gauge theories, the large number of degrees of freedom required for the dual theory to have a description in terms of classical gravity is achieved by considering a large number of colors, $N_c\to\infty$. At first sight, one would think that the gauge theory in this limit bears little resemblance to our case of main interest, $N_c=3$. But one would be wrong \cite{teper,panero}.

We have mentioned already that the gravitational dual of a non-Abelian gauge theory is a string theory (the content of which will be explained in the following section). This fact that had been variously anticipated since the early 70s. One reason is that the phenomenon of confinement for the strong interaction leads to the notion of the QCD string: a flux tube of the gluonic field (visible in lattice simulations), that for some purposes can be approximated as a one-dimensional object, and whose vibrational modes can be used to model aspects of highly-excited states in the hadronic spectrum (e.g., \cite{collins,teper,brandtmeineri}).
In this case, the physical operators $O(x)$ that we have discussed above are the composite operators that create hadrons.

Another reason is that the Feynman diagram expansion for a non-Abelian gauge theory with a large number of colors ($N_c\gg 1$), can be naturally reorganized as a series in powers of $1/N_c$, revealing a tantalizing parallel \cite{thooftlargen,mateosqcd,magoo} with the perturbative expansion of a theory of strings with coupling\footnote{For the case of 2-dimensional QCD, this parallel was shown in 1993 to follow indeed from a complete equivalence \cite{grosstaylor}.} $\propto 1/N_c$.
In the language of confining theories such as QCD, the physical statement is that, even if the partons themselves are strongly coupled within a given hadron, interactions among hadrons are suppressed by powers of $1/N_c$.  This suppression \cite{thooftlargen,wittenlargen} is in fact present even in nonconfining theories, and is consistent with our previous statement that the fields $\phi(x,z)$ of the gravitational description (dual to $O(x)$) become free as $N_c\to\infty$.   An important point to keep in mind is that, in a theory with many colors, the parameter that controls the validity of the Feynman diagram series is not the Yang-Mills coupling per se, $\gym^{2}$ (which figures in\footnote{Notice that we avoid the subindex `$s$' usually employed for the coupling associated with the $s$trong interaction, because the same letter is traditionally used for the $s$tring coupling.} $\alpha_{\mbox{\tiny YM}}\equiv\gym^{2}/4\pi$), but the combination $\lambda\equiv \gym^2 N_c$, known as the 't~Hooft coupling. The former gives the amplitude for emission of a gluon of a specific color, and the latter, the amplitude for  emission of a gluon of any of the many available colors, which is what really matters. So when we talk about strongly-coupled gauge theories, we mean $\lambda\gg 1$.

Historically, string theory itself in fact originated
as an attempt to understand hadrons in terms of one-dimensional constituents. This effort was largely abandoned when QCD became consolidated, but with the discovery of the holographic correspondence some 25 years later, we came to realize that it was not at all misguided. A myriad of developments in the intervening years brought the connection between gauge fields and strings into proper perspective: now we understand that the string theory in question lives on a different, higher-dimensional spacetime, and that it inevitably incorporates gravity.

\section{Strings from Gauge Fields, a Concrete Example}\label{concretesec}

In the previous section we got a general sense of what gauge-gravity duality asserts. Now we turn to a concrete example of a pair of theories identified under this duality, the example that was first discovered and is best understood.

\subsection{A cousin of QCD}\label{cousinsubsec}

On the field theory side, we start with QCD, with gauge group $SU(N_c)$ instead of $SU(3)$. Next, we eliminate the quarks, to be left only with the gluonic field $A_{\mu}(x)$ of pure Yang-Mills (YM) theory. Finally, we introduce a different set of matter fields: six real massless scalars $\Phi^S(x)$ ($S=1,\ldots,6$), and four Weyl fermions $\psi^{F}(x)$ ($F=1,\ldots,4$). Unlike the quarks, which were vectors with $N_c$ entries, in the fundamental representation of $SU(N_c)$, the new matter fields are chosen to be, just like the gauge field, $N_c\times N_c$ matrices, in the adjoint representation of $SU(N_c)$.

With carefully synchronized three- and four-point interactions associated with the single coupling $\gym$ that we started with (and constrained of course by gauge invariance), the resulting theory is known as $\cN=4$ or maximally supersymmetric Yang-Mills (MSYM) with gauge group $SU(N_c)$. `$\cN=4$' because, unlike the variety of supersymmetric theories considered useful for beyond-the-Standard-Model phenomenology, which have a single way of rotating a boson into a fermion or viceversa ($\cN=1$), this theory has four independent such rotations (whose four associated generators are Weyl spinors, and are known as supercharges). These symmetries in fact relate all of the fields with one another ($A_{\mu},\Phi^S,\psi^F$ form a single multiplet under supersymmetry), and are the reason why we had to choose specific numbers of each type of matter field, and why these had to be massless $N_c\times N_c$ matrix fields. They also completely determine the interactions. In a four-dimensional theory, $\cN=4$ is the largest amount of supersymmetry that one can have without involving gravity, which explains the phrase `maximally supersymmetric'. Unlike what happens in QCD or YM, in MSYM cancellations between fermion and boson loops imply that the coupling $\gym$ does not run with energy (the beta function vanishes). This means that the theory is scale-invariant: it is a CFT.

The six real scalar fields $\Phi^S(x)$ enter the theory on the same footing, so MSYM is symmetric under the global $SO(6)$ that rotates among them. The same can be said about the $SU(4)$ that rotates among the four complex $\psi^{F}(x)$. But these are not independent symmetries: just like  $SU(2)$ is the covering group of $SO(3)$, $SU(4)$ is the covering group of $SO(6)$, and in MSYM there are Yukawa-like couplings between the scalars and the fermions that preserve only one copy of $SO(6)\simeq SU(4)$. This is then the overall global internal symmetry of the theory. (It rotates the four supercharges among themselves, and because of this it is referred to as an `R-symmetry'.)

\subsection{A string theory}\label{stringsubsec}

On the gravity side, we have a string theory \cite{bbs},
whose basic excitations are ostensibly
one-dimensional strings with a characteristic length scale $l_s$, instead of zero-dimensional particles. By strings we mean for now closed strings, without endpoints. Different modes of vibration of these strings have the properties of distinct types of particles. Recall now that the central objects of study for a field theorist are not the particles but the fields themselves, with the former being just small fluctuations of the latter. In exactly the same way, the central object of study for a string theorist are not the strings, but some yet to be fully understood entity that for our purposes here can be referred to as the string field, whose small fluctuations are the objects we call strings. So in determining the modes of vibration of these strings, we are in fact considering different types of excitation of the string field.

Crucially, among these modes we invariably find one with all the properties of the graviton, the feeblest type of gravitational wave. This tells us that the string field incorporates spacetime itself as a dynamical object. Taking into account all the other string modes, each resembling a specific type of particle, we conclude that, at one level of understanding, the string field is a repackaging of spacetime together with an infinite tower of conventional fields. Innocent as this repackaging may sound, it in fact amounts to a drastic generalization of the notion of spacetime, with consequences that include the absence of UV divergences and dazzling equivalences between spaces with radically different sizes \cite{tduality}, or even with different topologies \cite{mirrorsymmetry}.
Among the surprises is the fact that, unlike in field theory or general relativity, the number of spacetime dimensions in a string theory cannot be chosen arbitrarily, but is fixed by requiring consistency at the quantum level (the absence of an anomaly).

 Situations that involve just a few strings interacting weakly can be described in terms of a perturbative expansion, using Feynman diagrams that are surfaces instead of graphs. The basic process is a string splitting into two, or the reverse, two strings fusing into one. Remarkably, this single interaction can reproduce an infinite variety of $n$-point interactions ($n\ge 3$) among the conventional fields associated with the different vibrational modes of the strings.  The parameter that determines the amplitude of the basic splitting process, and consequently controls the perturbative expansion, is known as the string coupling, $g_s$. Interestingly, its value is set by the environment: it is related to the expectation value of a massless scalar field $\varphi$ known as the dilaton, associated with a particular mode of vibration of the strings. When in the 90s an intricate web of string dualities gave us access for the first time to the regime where $\gs\gg 1$, we learned that, at a deeper level, strings are just an approximate description of more basic entities. All the known string theories are then subsumed into a single underlying framework known as M theory, whose complete definition is still unknown, but is not needed for our purposes here.

 The infinite set of conventional fields encompassed by the string field have masses proportional to the string scale $1/l_s$, so for energies $E\ll 1/l_s$ we need only consider the finite set of fields that are massless. String theory is then well approximated by a low-energy effective field theory that includes Einstein gravity. Successive corrections to this description involve terms with larger numbers of derivatives of the massless fields, multiplied, for dimensional reasons, by appropriate powers of $l_s$. In string theory we thus typically compute quantities in a double expansion: the series controlled by the coupling $g_s$, and the series controlled by the ratio of the string length $l_s$ to the radius of curvature of the background (spacetime plus other massless fields) under consideration (often rewritten in terms of $\alpha'\equiv\ls^2$). The quantumness and the stringiness of the string field depend respectively on how much $g_s$ and $l_s$ differ from zero.

 Whereas the perturbative excitations of the string field are strings, its nonperturbative excitations include extended, $p$-dimensional dynamical objects for various values of $p$, known collectively as branes. They are directly analogous to solitons of field theories; branes \emph{are} the solitons of string theory. Particularly important among these are D-branes, characterized by the fact that their tension is proportional to $1/g_s$, which is very large at weak coupling, but still smaller than that of other types of solitonic branes (whose tensions scale as $1/g^2_s$). We refer to these branes as D$p$-branes when we want to specify that they are extended in $p$ spatial dimensions. Small fluctuations of D$p$-branes are strings again, but now we mean open strings, whose endpoints can only slide along the $p$-dimensional volume of the D-brane. Just like we should think of closed strings as filaments obtained by exciting the stringy generalization of spacetime, open strings are in essence the filaments one obtains upon exciting the specific type of solitonic object that we call a D-brane. Modes of vibrations of these open strings, which are then modes of vibration of the D-brane itself, resemble specific types of particles, which can be interpreted as small fluctuations of fields living on the D-brane. These include massless scalar fields $\Phi^S$ related to displacements of the brane along the directions transverse to it, as well as a gauge field $A_{\mu}$.

 A key property of D-branes is that, if we form a stack of $N$ of them, all of the fields they give rise to become $N\times N$ matrices, because the open strings that describe excitations of the stack can begin and end on any one of the D-branes. This is true in particular for the gauge field, so the effective theory we obtain at energies $E\ll 1/l_s$ is a non-Abelian gauge theory. D-branes thus naturally make contact with this all-important structure of theoretical physics, and, if the ultimate goal of traditional string phenomenology is ever attained, it might well be the case that the Standard Model is reproduced solely with open strings, which would mean that we are nothing but excitations of a stack of D-branes! It is important however to emphasize that the application of string theory to holography is completely orthogonal to, and much less ambitious than, the search for a theory of everything.

 Returning to our general discussion, if we consider a stack with a large number of D-branes, we would expect it to substantially deform spacetime. And indeed, there exist solitonic solutions of the equations of motion of the fields arising from massless closed string modes, that have the appropriate mass and charge to be understood as an alternative description of a stack of D$p$-branes. These are known as black $p$-branes, generalizations of the concept of a charged black hole. As we will review momentarily, the relation between D-branes and black branes was precisely the starting point for Maldacena's derivation of the correspondence.

 The specific gravitational theory that we want to consider for now is known as IIB superstring theory (IIBST), where the `II' is meant to be a roman two. This theory is supersymmetric and lives in ten dimensions. Its massless sector includes the graviton $g_{MN}$ ($M,N=0,\ldots,9$), the dilaton $\varphi$, a second scalar $C$ known as the axion, and three gauge potentials $B_{MN},C_{MN},C_{MNPQ}$, analogous to the electromagnetic potential but with more indices. The low-energy effective field theory describing the dynamics of these fields and their fermionic superpartners is known as IIB supergravity (IIBSUGRA),
 and includes Einstein gravity, with Planck length $\lP\propto \gs^{1/4}\ls$. The solutions to the IIBSUGRA equations of motion identify backgrounds on which one can consistently define IIB string theory. The nonperturbative spectrum of IIBST includes D$p$-branes with  $p=1,3,5,7,9$ (and $-1$, which describes an object localized in time as well as space, an instanton).
 It also includes the so-called NS5-brane, a five-dimensional object whose tension scales as $1/\gs^2$. Whereas the strings are electrically charged under $B_{MN}$, the NS5 is magnetically charged under this same gauge field: it is a non-pointlike analog of a magnetic monopole. D-branes, on the other hand, are electrically or magnetically charged under the $C_{M\ldots}$ gauge fields.

 \subsection{Maldacena's derivation of AdS/CFT}\label{maldacenasubsec}

 Maldacena's argument \cite{malda} focused on a stack of $N_c$ D3-branes placed on flat ten-dimensional spacetime. This system has an alternative description in terms of a black threebrane, a solitonic background, translationally invariant along three directions, where only two fields are nontrivially excited. One is the metric $g_{MN}$, which in the transverse directions is akin to that of a charged black hole,\footnote{We mean here specifically a black hole that is extremal, i.e., that has the minimal possible mass for the given charge.} with a throat that connects a horizon at radial position $u=0$ to an asymptotically flat region at $u\to\infty$. The other excited field is the field strength $F_{MNPQR}$ for the electric-like potential $C_{MNPQ}$, which we will abbreviate as $F_{(5)}$. Using Gauss's law, the flux of this field strength through the five-dimensional sphere that surrounds the threebrane yields the total charge of the system, which we set to $N_c$ units to match the charge of the D-brane stack. Once we have done this, one indication that these two systems are in fact the same is the fact that the mass of the threebrane (inferred from the distortion of the metric in the distant, asymptotically flat region) coincides with the total mass of the D3-branes. Since it is $F_{(5)}$ that provides the stress-energy needed to curve the metric, there is a direct relation between $N_c$ and the radius of curvature of the geometry $L$, which takes the form $L^4=4\pi\gs N_c\ls^4$.

 Building on previous work, particularly by Klebanov and collaborators, what Maldacena noticed is that, starting with these two alternative descriptions and considering a limit of extremely low energies ($E\ll 1/\ls$ \emph{and} $E\ll 1/L$), we find an equivalence that is much more remarkable. On the D3-brane side we are left with MSYM on four-dimensional Minkowski spacetime, which is the non-Abelian gauge theory that captures the ultra-low energy dynamics of the D3-brane stack. On the black threebrane side, an extreme redshift effect implies that we are left still with the full IIBST, but restricted to the immediate vicinity of the horizon, where the metric simplifies and reduces to the product of AdS$_5$ and a five-dimensional sphere, S$^5$ (i.e., at each point in AdS we find five additional dimensions that form a sphere), with equal radii of curvature $L$ for both factors. The curvature is still supported by $N_c$ units of flux of $F_{(5)}$  through the five-sphere. Saying that we consider IIB string theory on this background means that (as in the more general discussion of Section~\ref{cftsubsec}) we allow dynamical excitations of it, small or large, in its interior, while the boundary conditions are kept fixed, so the ten-dimensional spacetime is always asymptotically AdS$_5\times$S$^5$.

 Assuming that the two initial descriptions for the system were truly equivalent, Maldacena was then led to the conclusion that the two descriptions obtained after his ultra-low energy limit had to be equivalent, too:\footnote{To avoid possible confusion, there is a point here that we should emphasize. Even though extremely low energies are required to \emph{derive} this statement of equivalence starting from a D3-brane stack (because it is only then that the physics simplifies in the manner described), in the resulting theories (after the massive simplfication) \emph{we are free to explore energies as large as we like}, without returning to the full D3 setup. This is particularly evident on the field theory side, where we are left with a CFT, which is clearly well-defined by itself at arbitrary energies. The duality associated with the D3 stack at energies that are low compared to the string scale but still away from the Maldacena limit is more elaborate, and has been explored in \cite{d3holo}-\cite{harmark} (see in particular Section 2 of \cite{coniholo}).}

 \vspace*{0.2cm}
 \hspace*{1cm}
\framebox{
 \begin{minipage}{2in}
\centering
MSYM on R$^{3,1}$ \\
 with gauge group $SU(N_c)$
 \end{minipage}
{\huge =}
 \begin{minipage}[c]{2in}
 \centering
IIBST on AdS$_5\times$S$^5$ \\
 with $N_c$ units of $F_{(5)}$ flux
 \end{minipage}
 ~.}
 \vspace*{0.2cm}

\noindent The claim is that, in spite of their many evident differences, these two theories are in fact just two different languages to describe the same physical system, and there is a dictionary that translates between them. The fact that we know the origin of the equivalence allows us to identify many of the entries of this dictionary.

\subsection{The dictionary, and some evidence}\label{dictionarysubsec}

The parameters of the two theories are related through
\begin{equation}\label{msymcouplings}
  \gym^2=4\pi\gs~,\qquad \lambda\equiv\gym^2 N_c = L^4/\ls^4~,
\end{equation}
which in turn imply that $N_c\propto L^4/\lP^4$. All evidence suggests that the equivalence between the two theories holds for any value of these parameters, but calculations on the IIBST side are under quantitative control only when the string coupling is small and the radius of curvature of the spacetime is large in string units. {}From (\ref{msymcouplings}) we see immediately that in the gauge theory this is the previously announced regime where the number of colors is large and the coupling is strong, $N_c\gg 1$, $\lambda\gg 1$.
The AdS$_5$ coordinates are transcribed into MSYM in the manner described in Section \ref{energysubsec}, with the energy scale $\mu$ of the renormalization group becoming a fifth spatial dimension. And the five-sphere that we encounter in the gravitational description indicates that other variables of the gauge theory have become geometrized too: specifically, angular position on S$^5$ refers to the direction in the abstract internal space on which the six scalar fields $\Phi^S$ (and the four spinor fields $\psi^F$) take values.

A first piece of evidence for the equivalence is the fact that the global symmetries of the two descriptions match exactly. As we mentioned in Section~\ref{cftsubsec}, the isometry group of AdS$_5$ agrees with the four-dimensional conformal group (remember that MSYM is a CFT). The isometry group of S$^5$ is $SO(6)\simeq SU(4)$, and agrees with the internal symmetry group that rotates among the $\Phi^S$ and, separately, among the $\psi^F$. Supersymmetries also match perfectly. Local symmetries do not agree: the $SU(N_c)$ of MSYM is nowhere to be found on the gravity side, and conversely, the ten-dimensional diffeos of IIBST are not present in the gauge theory side. But we were already prepared for this mismatch: as explained in Section~\ref{basicsec}, it poses no problem because local symmetries are nothing but redundancies of each description.

The duality is only obligated to translate physical quantities. Among these are local gauge-invariant operators $O(x)$ in MSYM, which can be interpreted as operators that create glueballs.\footnote{In truth, scale invariance precludes the theory from having discrete particle states when formulated on R$^{3,1}$ (see, e.g., \cite{georgi}), but it is convenient to use this intuitive language nonetheless, which will be completely appropriate for the confining examples discussed in the next section. A discrete spectrum is also obtained if we consider MSYM not on Minkowski but on S$^3\times$R \cite{horowitzooguri}.} These operators are composite, built by starting with the basic matrix fields $F_{\mu\nu}, \Phi^S, \psi^F$ and covariant derivatives thereof, multiplying them in sequence, and then taking the trace to do away with the color indices. We can also take products of such traces.  These $O(x)$ were anticipated in Section \ref{cftsubsec} to be dual to fields $\phi(x,z)$ on AdS$_5$.

In IIBST we have, to begin with, the fields associated to the different vibrational modes of the strings, which are defined on the full ten-dimensional AdS$_5\times$S$^5$ spacetime. To interpret them as fields on AdS, we expand in terms of spherical harmonics on the five-sphere, $\phi(x^M)=\sum_I \phi_I(x^{\mu},z)Y_I(\theta^a)$, where $I$ (for `internal') represents jointly all the relevant subindices of the harmonics $Y$,
 and we have suppressed possible spacetime indices of $\phi$, which lead to further multiplicity. In this way we obtain, from each ten-dimensional field, one or more infinite towers of Kaluza-Klein modes $\phi_I(x,z)$ on AdS$_5$, in specific representations of the $SO(6)$ group, precisely as we need to have a chance of matching the MSYM operators, which indeed belong to such representations and can thus be denoted $O_I(x)$. Notice that, unlike what happens in traditional string phenomenology, where one compactifies dimensions to hide them, here the compact space, the sphere, is not small. Its radius $L$ is equal to the radius of curvature of AdS, so if we start with a field $\phi$ that is massless from the ten-dimensional perspective, its Kaluza-Klein progeny on AdS$_5$ will have masses $\propto 1/L$, and all of these modes should be kept.

As is partly indicated by the subindex $I$, each $O_I(x)$ belongs to a family of operators related by the global symmetries (i.e., a superconformal multiplet). Since the symmetries of the two theories agree, the $\phi_I(x,z)$ will automatically belong to such families as well. What is not at all guaranteed, however, is that in each theory one will find precisely the same number of each possible type of family. These are classified as usual by the Casimirs of the symmetry group, which have a different interpretation on each side of the duality. As a result, it happens that the scaling dimension $\Delta$ of the $O_I$ is related to the (five-dimensional) mass $m$ of the corresponding $\phi_I$. In the case of scalar operators, for example, the relation takes the form $m^2 L^2=\Delta(\Delta-4)$.

Given what we know about the spectrum of IIBST at $g_s,\ls/L\ll 1$, we then have an infinite set of specific predictions for the scaling dimensions and other quantum numbers of the operator families that we should find in strongly-coupled large-$N_c$ MSYM. Of course, the fact that $\lambda\gg 1$ makes these predictions difficult to test in general. But  certain kinds of operators have special properties that allow their dimension to be determined at large $\lambda$, allowing a direct comparison to the gravity predictions. This is the case for all $\phi_I$ arising from ten-dimensional IIB supergravity modes (arising from the lightest excitations of the strings), whose decomposition into S$^5$ harmonics was worked out long ago \cite{gm,krvn}, and later shown \cite{w} to agree perfectly with the available families of operators in MSYM. For instance, the graviton field on AdS, $g_{mn}(x,z)$, is found to correspond to the stress-energy tensor $T_{\mu\nu}(x)$, as we had advertised in Section \ref{genericsubsec}. Another example is the dilaton $\varphi(x,z)$, which corresponds to the Lagrangian density $(1/2\gym^2)\tr[F^2(x)+\ldots]$.

Notice that this successful matching between a class of objects in both theories means in particular that MSYM secretly contains gravitons and all other ten-dimensional supergravity modes! Recalling that as a consequence of (\ref{msymcouplings}) the quotient that controls the strength of the IIBSUGRA interactions is $\lP/L\propto 1/N_c^{1/4}$, we see that on both sides of the duality these objects are free in the strict $N_c\to\infty$ limit, and weakly coupled for $N_c$ large but finite. Going beyond supergravity, an even more impressive matching has been shown to exist for full towers of stringy modes with large quantum numbers \cite{bmn}-\cite{hofmanmalda},
signaling that MSYM, in spite of its non-confining nature, secretly contains strings!\footnote{And there is more: beyond accounting for supergravity or stringy modes that represent small fluctuations of AdS$_5\times$S$^5$, the gauge theory also has states that describe \emph{large} (but normalizable) deformations of the background. See, e.g., \cite{llm} and Section \ref{tempsubsec}.}

Having achieved highly non-trivial agreements between large classes of glueballs in MSYM and closed string modes in IIBST,
the next step is to consider whether these objects interact in the same manner in both descriptions. In the gauge theory, this information is contained in correlators of the glueball operators $O_I(x)$. These can be summarized in terms of a partition function
$Z_{\mbox{\tiny MYSM}}[J_I]\equiv\int\mathcal{D}(A,\Phi,\psi)\exp[iS_{\mbox{\tiny MSYM}}+i\int d^4 x J_I(x)O_I(x)]$,
which includes a source $J_I(x)$ for each operator of interest. To compute correlators in MSYM, one functionally differentiates with respect to $J_I$ and then sets all sources to zero; but in $Z[J_I]$ itself, when $J_I\neq 0$ we are dealing with a theory that differs from MSYM by the addition of terms linear in the $O_I$. According to what we described in the Section~\ref{genericsubsec}, on the gravity side this should correspond to turning on the non-normalizable part of $\phi_I$, i.e., changing the boundary conditions. If this field has mass $m$, the leading radial dependence near $z=0$ allowed by the equation of motion is $\phi_I(x,z)=z^{4-\Delta}\phi_I^{(0)}(x)$ (with $\Delta$ the dimension of $O_I$), so the boundary conditions are set by specifying $\phi_I^{(0)}(x)$. And indeed, we had emphasized in Sections~\ref{cftsubsec}-\ref{genericsubsec} that to define a theory on AdS, because of the asymptotic structure of (\ref{adsmetric}) at spatial infinity, boundary conditions must be specified. Once this is done, one can in principle compute the string theory partition, which morally takes the form
$Z_{\mbox{\tiny IIBST}}[\phi_I^{(0)}]\equiv\int\mathcal{D}(g,\varphi,\ldots)\exp[iS_{\mbox{\tiny IIBST}}]$
(notice this does not include sources in the bulk, because there are no local observables in a gravitational theory).

A crucial entry in the dictionary of the duality equates the partition functions of the two theories at equal values of their corresponding functional arguments \cite{gkp,w},
\begin{equation}\label{zz}
Z_{\mbox{\tiny MYSM}}[J_I]=Z_{\mbox{\tiny IIBST}}[\phi_I^{(0)}]\qquad\mbox{for}\quad J_I(x)=\phi_I^{(0)}(x)~.
\end{equation}
In practice, we can only compute the right-hand side in a regime where it drastically simplifies. For $N_c\gg 1$ (meaning $g_s\ll 1$), the path integral can be treated in a saddle-point approximation, and for $\lambda\gg 1$ (meaning $L\gg l_s$), string theory can be approximated by supergravity, $S_{\mbox{\tiny IIBST}}\simeq S_{\mbox{\tiny IIBSUGRA}}$. We are then left with a remarkable recipe for obtaining correlators in the strongly-coupled gauge theory in terms of the classical supergravity action evaluated on the solution to the equations of motion, $Z_{\mbox{\tiny MYSM}}[J_I]\simeq\exp[iS^{\mbox{\scriptsize on-shell}}_{\mbox{\tiny IIBSUGRA}}]$.

Refs.~\cite{gkp,w} and most of the initial calculations of correlators were carried out in Euclidean signature. In the Lorentzian case one has additional liberties \cite{vijay,hs}, and a complete prescription was given in \cite{svr}. The recipe for correlators allows many further checks of the correspondence, as well as a large number of interesting predictions. At $N_c\to\infty$, both MSYM and IIBST are known to become integrable, and this has led to very
impressive matches of both the spectrum and the interactions for \emph{arbitrary} values of the coupling $\lambda$ \cite{beisert}.

\subsection{Holographic renormalization and the RG}

When computing correlators using the relation $Z_{\mbox{\tiny MYSM}}[J_I]\simeq\exp[iS^{\mbox{\scriptsize on-shell}}_{\mbox{\tiny IIBSUGRA}}]$, one encounters divergences in $S^{\mbox{\scriptsize on-shell}}_{\mbox{\tiny IIBSUGRA}}$, due to the fact that the volume of the AdS spacetime (\ref{adsmetric}) (or its non-conformal generalizations) diverges at $z\to 0$. This is the gravity counterpart of the familiar UV divergences of quantum field theories, and are dealt with in an analogous manner. First, one regularizes by choosing a small parameter $\epsilon$ and temporarily removing the offending $0\le z<\epsilon$ region of the geometry. This is a sharp IR cutoff in the gravity language, but corresponds to a UV cutoff in the field theory. For MSYM or other examples of holography involving non-Abelian gauge theories, this simple procedure achieves something previously unheard of in the field theory language: it implements a UV cutoff that is at the same time gauge-invariant (unlike, say, a brute-force cutoff), non-perturbative (unlike dimensional regularization) and Lorentz-invariant (unlike the lattice).

Next, without modifying the physics in the $z>\epsilon$ region, one makes sure that the theory is properly defined in the $\epsilon\to 0$ limit, by adding to the bulk action boundary terms, defined on the cutoff surface $z=\epsilon$.
Because of the analogy with the field theory, these additional contributions to the action are known as counterterms, and the process of introducing them is known as holographic renormalization. Good reviews are found in \cite{skenderis,papadimitrious,svr}.

As usual, there is some freedom in choosing the finite parts of the coefficients in these counterterms, corresponding to the expected liberty of using different renormalization schemes. This naturally leads to the holographic implementation of the renormalization group \cite{dbvv,fukuma,kln} \`a la Callan-Symanzik, where the values of the bulk fields $\phi_I(x,z)$ at different radial depths translates into the evolution of the couplings $g_I(\mu)$ for the various operators $O_I(x)$. This explicitly realizes the geometric picture that we delineated in Section~\ref{basicsec}. In fact, what we get is a \emph{local} generalization of the RG, where the couplings are allowed to vary with $x$, as in \cite{osborn}.
One can also implement the renormalization group \`a la Wilson, by maintaining the UV cutoff $\epsilon$ in place, and studying the evolution of the boundary action as $\epsilon$ is increased \cite{hpwilsonian,flrwilsonian,sslee,bkm}. In the presence of an explicit UV cutoff, one can additionally envision generalizations of the correspondence to the case of field theories that are not UV-complete \cite{bklswilsonian,surfacestate}.

\subsection{External quarks, mesons, baryons and radiation}\label{quarksubsec}

Another time-honored way to explore the dynamics of any gauge theory is to couple it to external (infinitely heavy, classical) charges, and study the response of the fields. In a non-Abelian theory, we can choose our color sources in any representation of the gauge group, with the most familiar case being of course quarks, i.e., sources in the fundamental representation of $SU(N_c)$. Going back to the stack of $N_c$ D3-branes whose low-energy dynamics gave rise to our MSYM=IIBST duality, one can reason that, just  like the $N_c\times N_c$ matrix (adjoint) fields of MSYM arose from open strings with both of their endpoints on the stack, the $N_c$-entry fundamental field that describes a quark should originate from an open string with only one of its endpoints on the stack, and the other endpoint away from it (on some other D-brane, which may or may not be a D3-brane). In the gravity description, where the D3-brane stack is replaced by the entire AdS$_5\times$S$^5$ spacetime, the counterpart of an isolated quark ends up being \cite{reyyee,maldawilson} a macroscopic string extending all across this geometry, with one endpoint at $z\to\infty$ and the other at $z=0$. The energy of such a string is infinite, as befits the dual of an infinitely massive quark. For $\lambda, N_c \gg 1$, this string behaves classically.

An antiquark corresponds to an anti-string, that is, a string with the opposite orientation, and therefore opposite charge under $B_{MN}$. When we have both a quark and an antiquark, the lowest energy configuration on the gravity side for the given boundary conditions is a U-shaped string with both of its endpoints on the boundary, which is then the dual of an external meson. An external baryon contains $N_c$ quarks, but we would seem to be out of luck, because in the gravity description there is a priori no way for $N_c$ strings ending with the same orientation at the AdS boundary to join together in the bulk (their other endpoints cannot just stick together). Two very specific properties of our IIBST setup save the day: the existence and properties of D5-branes, combined with the $N_c$ units of the $F_{(5)}$ flux, allow a D5-brane wrapped on the S$^5$ to act as the glue that holds together precisely $N_c$ strings \cite{wittenbaryon,go}. In so doing, the D5 gets pulled by the strings and is ultimately deformed into a tube that reaches all the way to the AdS boundary \cite{baryon}. So a baryon is dual to a D5-brane! More generally, the D3-branes and D5-branes present in IIBST have precisely the right properties \cite{df,yamaguchi,gp} to account for $k$-quarks in arbitrary representations of $SU(N_c)$ (which are in one-to-one correspondence with Young tableaux). This remarkable agreement serves to illustrate how crucial have been the advances in our understanding since the discovery of string theory in the early 70s, for it was only in the 90s that we learned about the presence of D-branes in the non-perturbative spectrum.

This assortment of color sources allows us to extract a large variety of quantities in the strongly-coupled gauge theory from rather simple, classical computations in the gravity language. In more detail, infinitely massive charges are added to the gauge theory by inserting Wilson loops (or lines) into the correlators. The non-local gauge-invariant operator $W_{\mathcal{R}}(C)\equiv \tr_{\mathcal R}[P\exp(i\int_C A\cdot dx+\ldots)]$ computes the phase accumulated by a color source in representation $\mathcal{R}$, when it follows a specified trajectory $C$ in spacetime and in the internal space. Among other uses, Wilson loops serve to determine the quark-antiquark potential, to diagnose phase transitions, and to define the cusp anomalous dimension, a quantity that has a number of important applications in QCD \cite{korchemsky}.
The fact that $k$-quarks in MSYM are dual to strings or D-branes in IIBST ends up leading to a concrete recipe for computing correlators with Wilson loops: in the gravity side, one must perform the path integral requiring as a boundary condition that an infinite string or D-brane ends at $z=0$, tracing the curve $C$ there.

For $N_c, \lambda\gg 1$ this again simplifies drastically, and we just need to extremize the appropriate classical action. E.g., since a quark ($\mathcal{R}=\square$) is dual to a string, the recipe in this case is \cite{reyyee,maldawilson} $\langle W_{\square}(C)\rangle \simeq\exp[iS^{\mbox{\scriptsize on-shell}}_{\mbox{\tiny string}}]$. One example is to take $C$ to be a rectangle of spatial width $\ell$ and extension $\mathcal{T}\to\infty$ along the time direction, from which one can infer the quark-antiquark potential, $\langle W_{\mathcal{\square}}(C)\rangle=\exp[iV_{q\bar{q}}(\ell)\mathcal{T}]$. A simple calculation \cite{reyyee,maldawilson} yields a Coulomb potential, exactly as expected because of the conformal invariance of MSYM, but with a coefficient $\propto\sqrt{\lambda}$ instead of the factor of $\lambda$ encountered at leading order in perturbation theory. This dependence on the coupling is in fact predicted by AdS/CFT for all Wilson loops in the fundamental representation, at large $N_c$ and $\lambda$. For a class of circular Wilson loops with specific types of charge under the MSYM scalar fields $\Phi^S$, a surprising trick known as localization \cite{pestun} allows one to compute from the field theory side the \emph{exact} result \cite{dg,esz,pestun} for any value of $N_c$ and $\lambda$, and in the appropriate regime it indeed agrees with the corresponding AdS/CFT prediction. Other quantities and representations have been addressed in a similar manner (e.g., \cite{nadavwilson}-\cite{leopzwilson}).

One can also compute correlators of a Wilson loops and any choice of local operators. For instance,
$\langle O(x)\rangle_q\equiv\langle W_{\mathcal{\square}}(C)O(x)\rangle/\langle W_{\mathcal{\square}}(C)\rangle$
maps out the profile of the
gluonic (and other) field(s) sourced by a quark. In the case of a static quark at the origin, this gives \cite{dkk,cg} the expected Coulombic result,
$\langle(1/2\gym^2)\tr(F^2(x)+\ldots)\rangle_q\propto\sqrt{\lambda}/|\vec{x}|^4$. A color-neutral source like a meson or a baryon is predicted instead \cite{cg} to yield a $1/|\vec{x}|^7$ falloff, suppressed with respect to the familiar $1/|\vec{x}|^6$ dipolar falloff. Interestingly, this predicted suppression was later understood to be an effect of the large $N_c$ limit, present even at weak coupling \cite{kmt}.

In the case of an accelerating quark, one can study the propagation of disturbances of the gluonic field \cite{cg,gubserdilaton}, encountering the $1/|\vec{x}|^2$ falloff characteristic of radiation \cite{mo,gubsergraviton1}, and making interesting inferences about the angular and temporal distribution. In spite of the strong coupling, beams of radiation can remain collimated instead of rapidly spreading out \cite{liusynchrotron,veronika2,beaming}. And even though the nonlinear character of the gluonic medium implies that signals are reradiated from all possible scales \cite{cg}, the net disturbance is found not to broaden in the radial/temporal direction \cite{iancu1,iancu2,trfsq,tmunu}.
Exact results have also been found in this context \cite{lmexact}.

An important lesson of these studies is that the quark itself (the pointlike source of color in the fundamental representation) is dual only to the endpoint of the string on the AdS boundary, whereas the body of the string encodes the gluonic field sourced by the quark. In other words, the latter is the direct analog of the QCD string, even though in the present, nonconfining setting, our color `flux tubes' have a natural tendency to spread out \cite{noline}, and can even encode gluonic radiation!

\subsection{Entanglement entropy}\label{entanglesubsec}

What we have discussed up to now is the traditional way to probe the quantum state of a field theory (e.g., the vacuum), by placing on it operators $O(x)$ or $W_{\mathcal R}(C)$. The information revealed in this manner depends both on the state and on the nature of the probe. Since the late 90s and early 2000s, an alternative approach has been intensely pursued, that gives access directly to the properties of the state without needing to specify a particular probe. This involves studying the pattern of spatial entanglement present in the state, which after all is the reason why correlators of operators inserted at spacelike-separated points do not vanish \cite{hastings}.

More concretely, given a state $\ket{\psi}$ and a spatial region $A$ in the field theory, one can construct a reduced density matrix by tracing over the degrees of freedom in the complementary region $A^{\mbox{\scriptsize c}}$: $\rho_A\equiv\tr_{A^{\mbox{\tiny c}}}(\rho)$, with $\rho\equiv\ket{\psi}\bra{\psi}$ (or a more general density matrix if the overall state is mixed).  The entanglement entropy is then defined as the von Neumann entropy $S_{\mbox{\scriptsize ent}}\equiv -\tr(\rho_A \ln\rho_A)$, and measures the amount of (classical and quantum) correlation that exists in the given state between the degrees of freedom in $A$ and those in $A^{\mbox{\scriptsize c}}$. This quantity has been found to provide a very useful handle on field theories, with connections to the renormalization group and other interesting properties \cite{cc,ch}, but is difficult to compute, even for free field theories.

An important development occurred in 2006, with the formulation of a very simple recipe to calculate the entanglement entropy at strong coupling, using AdS/CFT. For static situations, the prescription is to draw the region $A$ on the AdS boundary, and identify the surface $\Sigma$ that hangs down into the bulk from the edge of $A$, can be continuously deformed to become $A$, and has the smallest possible area. It was proposed in \cite{rt} and later proved in \cite{lm} that, in the region where the gravitational dual involves classical Einstein gravity, the entanglement entropy is then given by
\begin{equation}\label{rt}
S_{\mbox{\scriptsize ent}}=\frac{\mbox{area}(\Sigma)}{4G_N}~,
\end{equation}
where $G_N$ is Newton's constant.
This formula, closely related to the famous black hole entropy formula of Bekenstein and Hawking \cite{bekenstein,hawking},
has made possible an enormous body of work \cite{takayanagi,rangamanitakayanagi}. Generalizations have been made to time-dependent situations \cite{hrt,dlr}, to higher-derivative gravities \cite{dong1,camps}, to Renyi entropies (which contain further information about the pattern of entanglement) \cite{dong2},  to `entwinement' (a notion of entanglement among gauge-variant degrees of freedom) \cite{bartekentwine,vijayentwine}, and to the first quantum ($1/N_c$) corrections \cite{flm,bdhm}.

The most remarkable aspect of (\ref{rt}) is that it establishes a direct link between entanglement, the quintessentially quantum feature, and classical geometry. This has led to the recognition of entanglement as the central ingredient for reconstructing the bulk spacetime, starting from the field theoretic language \cite{rangamanitakayanagi,vr}-\cite{harlowjerusalem}.
An important related advance is a very recent reformulation of the AdS/CFT dictionary between field theory operators $O(x)$ and the corresponding bulk fields $\phi(x,z)$, in a manner that is manifestly invariant under bulk diffeos, and sensitive to depth along the radial direction \cite{stereobartek,stereomyers,stereomonica}.

\subsection{MSYM vs.~QCD}\label{comparisonsubsec}

Having learned that a particular non-Abelian gauge theory, MSYM, is completely equivalent to a string theory, and that upon translating to the gravitational language we gain access to the strongly-coupled regime of the field theory, it is natural to ask how closely related MSYM is to its cousin of main interest for us in this review, QCD. If we compare these two field theories for small excitations around the vacuum, the answer is definitely not much. QCD is scale-dependent, non-supersymmetric, and confining, as a result of which it has a massive spectrum and a linear quark-antiquark potential. MSYM is conformal, highly supersymmetric and (on R$^{3,1}$) does not confine, as a result of which it has no mass gap (and strictly speaking, no asymptotic states at all) and a Coulombic quark-antiquark potential. Essentially the only point of contact is that both theories have the same gluonic amplitudes at tree level, and some colored matter fields. Near the vacuum, then, MSYM is for most purposes not a useful toy model of QCD.

On the other hand, if we compare the two theories at a finite temperature $T$, larger than the deconfinement temperature in QCD ($T_{\mbox{\scriptsize c}}\sim\Lambda_{\mbox{\scriptsize QCD}}$, about a trillion Kelvin), the situation improves considerably. The system under consideration in QCD will then be the quark-gluon plasma (QGP) \cite{mateos}, and in MSYM we will have a thermal plasma of gluons and adjoint fermions and scalars. The supersymmetry of MSYM is no longer an issue, because it is broken by the finite temperature (since the thermal distributions for bosons and fermions are different). In both theories, the potential is Coulombic at short distances and screened by the plasma at large distances. In MSYM, the temperature provides the only available scale, so the $T$-dependence of thermodynamic quantities is determined by dimensional analysis. In QCD one also has the scale associated with the running of the coupling, so a priori these same quantities could behave in a more complicated manner. Lattice results \cite{karsch,petreczky}, however,
show that in a window where $T$ is just a few times higher than $T_{\mbox{\scriptsize c}}$, the behavior is in fact controlled predominantly by the temperature alone. In short, at finite temperature MSYM does serve as a toy model of certain aspects of QCD, even if it is a rudimentary one.

\subsection{Heating and stirring things up}\label{tempsubsec}

To turn on a finite temperature $T$ in MSYM, we can run through Maldacena's derivation as described in Section \ref{maldacenasubsec}, but starting with a stack of D3-branes with an added spatially-uniform energy density $\mathcal{E}$ (carried by a fluid of open strings), noting that its alternative description is a black threebrane with finite temperature. Or we can stay within our now familiar MSYM=IIBST framework, and simply use its dictionary to work out the metric dual to a field theory state where the expectation value of the stress-energy tensor corresponds to a static, relativistic fluid in a CFT, $\expec{T_{\mu\nu}(x)}=\mbox{diag}(\mathcal{E},p,p,p)$ with pressure $p=\mathcal{E}/3$ on account of conformal invariance (which implies that $\eta^{\mu\nu}T_{\mu\nu}=0$). Either way, one finds \cite{malda} that the relevant geometry is that of a neutral (Schwarzschild) black hole in AdS$_5$ (more precisely, an asymptotically-AdS black threebrane),
\begin{equation}\label{schwadsmetric}
 ds^2=\frac{L^2}{z^2}\left[-\left(1-\frac{z^4}{z_h^4}\right)dt^2+d\vec{x}^{\,2}+\frac{dz^2}{\left(1-\frac{z^4}{z_h^4}\right)}\right]~, \qquad z_h=\frac{1}{\pi T}~,
\end{equation}
times the ever-present five-sphere. $z=z_h$ here marks the location of the black hole's event horizon, and $T$ is its Hawking temperature \cite{hawking}, which is directly identified with the field theory temperature.

Using the black hole metric (\ref{schwadsmetric}), we can explore a large number of properties of the gluon (plus adjoint matter) plasma that we have in large-$N_c$ MSYM at finite temperature. The first famous result was the prediction that the entropy density $s$ of the plasma at infinite coupling is simply reduced by a factor of $0.75$ from its value at zero coupling \cite{gkpentropy}, which is notable because the lattice results \cite{karsch,petreczky} for the real-world QGP (in QCD with $N_c=3$) show an almost identical reduction, by a factor of $\sim 0.8$.

One can also access dynamical quantities of the strongly-coupled soup, moving from thermodynamics to hydrodynamics \cite{pss}. In particular, the linear-response approach to physics out of equilibrium relates the shear viscosity $\eta$ of the plasma to the low-frequency limit of a retarded two-point correlator for the stress-energy tensor. On the gravity side, this translates into a computation of the absorption cross-section for gravitons by the black hole. This seemingly ludicrous connection leads to the celebrated prediction \cite{kss,ssreview} $\eta/s=1/4\pi$ for the ratio of viscosity to entropy density. This result was shown to be universal, in the sense that it applies to all strongly-coupled thermal plasmas with a gravity dual \cite{bluniversal,cremonini}, and remarkably, it is quite close to the experimentally-estimated ratio for the QGP \cite{song}.\footnote{To avoid lingering confusion, we should also emphasize that the independent conjecture \cite{kss} that $1/4\pi$ provides a lower bound for $\eta/s$ in all relativistic field theories is \emph{not} a prediction of AdS/CFT, and moreover, turns out to be incorrect \cite{cremonini}.}

On the heels of this remarkable semi-quantitative agreement with experiment, a tremendous amount of phenomenologically-motivated work was carried out, to extract a myriad of other properties of strongly-coupled thermal plasmas, including their response to heavy partons traversing them. Good reviews of varying extension can be found in \cite{karchrecent,dwgrt,gkpedestrian,mateos}.

The linear-response analysis of the field theory corresponds in the bulk to studying linearized gravity (+dilaton+etc.) on the static black hole (\ref{schwadsmetric}). In 2007, a remarkable theoretical development made it possible to examine out-of-equilibrium physics way beyond this linear approximation. Indeed, by systematically working order by order in the derivative expansion naturally associated with the hydrodynamic regime of the CFT, one can obtain a complete match between the relativistic Navier-Stokes equations for the fluid and the full \emph{nonlinear} Einstein equations that control the evolution of a dynamical black hole in AdS.
This translation is known as the fluid-gravity correspondence \cite{fluidgravity}, and has been nicely reviewed in \cite{hmrreview,hrreview}. Its discovery amounts to a proof of the AdS/CFT correspondence, for the restricted class of states where $\expec{T_{\mu\nu}(x)}$ has an arbitrarily large amplitude but a small rate of variation in spacetime.

We should emphasize, however, that AdS/CFT does not match \emph{only} onto hydrodynamics. In the linearized approach it gives us access to a complementary regime, where the amplitude of $\expec{T_{\mu\nu}(x)}$ above its equilibrium value is small, but its rate of spacetime variation can be arbitrarily large. And its range of applicability extends much farther still, if one is willing to perform numerical calculations. This has been exploited in particular to model the evolution of the RHIC/LHC fireball from a set of initial conditions all the way to the hydrodynamic regime \cite{chesler,heller,dwgrt,mateos}.

\section{Other Examples, and Some Approaches to QCD}\label{othersec}

In view of the differences between MSYM and QCD discussed in Section~\ref{comparisonsubsec}, it is natural to ask next if there exist other examples of gauge-gravity duality involving field theories that are closer cousins of QCD. This question has been approached from two different perspectives.

The ideal route is to have a derivation originating from constructions in string (or M) theory, and leading to the identification of a specific field theory and a specific gravitational theory that are dual to one another. This is known as the `top-down' approach, and provides us with a firm basis for believing that a duality exists, as well as with some degree of control over the dictionary that implements it. A large catalog of examples has been worked out.
The one disadvantage of this approach is that the range of theories that we obtain is limited by the demand that we understand the origin of the duality, so we do not usually get quantitative access to the field theories we would most like to study.

The other, complementary route is to build on the basic intuition of holography described in Section \ref{basicsec} to postulate a gravity dual with ingredients chosen by hand to resemble the features of a desired field theory. This is known as the `bottom-up' approach, and is intrinsically phenomenological. One performs computations in the gravitational description of a field theory whose exact nature, and even existence, is uncertain. In exchange for this, one has more leeway to try to come close to the desired physics.

{}From a theoretical standpoint, the top-down approach is of course more satisfactory, because it allows calculations from first principles in a controlled setting, in theories that are of interest in their own right. But when extracting lessons for real-world systems, it must be acknowledged that one is computing in the wrong theory, and extrapolation to the correct theory is not a controlled approximation. So at this point the approach becomes phenomenological, and part of the distinction with the bottom-up approach disappears.
In this section we will discuss both routes, highlighting one example of each.

\subsection{Two roads to top-down constructions}\label{topdownsubsec}

Within the top-down approach, there are two ways to obtain new examples of the correspondence. One is to start with a known example and turn on non-normalizable modes of the available fields. We mentioned already in Section~\ref{genericsubsec} that this changes the theory, a point that we saw more concretely in Section~\ref{dictionarysubsec}, in terms of the recipe for the partion function. For the most part, one thinks of $Z_{\mbox{\scriptsize QFT}}[J_I]$ as a tool to compute correlators in a given theory, so after functionally differentiating with respect to the sources $J_I$, one sets them to zero. But we can also leave some of the sources on, in which case we are computing with a Lagrangian deformed by the addition\footnote{Of course, we would then usually want to take the sources $J_I=$constant at the end of the calculations, so that they play the role of spacetime-independent couplings.} of the corresponding operators $O_I$.

 On the gravity side, this translates into modifying the boundary conditions for the dual fields $\phi_I$. We can start, for instance, with MSYM and turn on masses for some of the scalar and fermion fields, thereby reducing the amount of supersymmetry (or eliminating it altogether), and inducing a renormalization group flow. Sometimes the flow can be consistently followed within the range of validity of classical IIBSUGRA, or even within a truncation of it to five dimensions. In other cases this range is surpassed and new ingredients of the full IIBST come into play, or we might lose quantitative control of the string description. Examples are known where the deforming operators are marginal ($\Delta=4$), meaning that the theory remains conformal \cite{luninmaldacena},
and where they are relevant ($\Delta<4$), with the flow
leading to an IR fixed point \cite{fgpw} or to confinement \cite{ps}, perfectly matching field theory expectations whenever they are available\footnote{A very basic agreement follows immediately from the near-boundary behavior that we described in the paragraph above (\ref{zz}). The non-normalizable mode for a field dual to an operator $O(x)$ has the radial dependence $\phi_I(x,z)=z^{4-\Delta}\phi_I^{(0)}(x)$ as $z\to 0$. If we use this operator to deform the CFT, then for $\Delta<4$ (a relevant deformation) the field switches off as we approach the AdS boundary, just as we expect from the field theory side, where we still have a fixed point in the UV. On the other hand, for $\Delta>4$ (an irrelevant deformation), the boundary condition blows up as $z\to 0$, meaning that the asymptotic structure is completely changed, which reflects the fact that we no longer have a UV fixed point.} \cite{ls,donagiwitten}.

A second way to arrive at new top-down constructions is to run through Maldacena's derivation starting from a different system. Instead of a stack of D3-branes we can consider other types of branes \cite{malda,wittenthermal}-\cite{abjm},
 again on flat spacetime or on other geometries \cite{kw}-\cite{bgmpz}.

\subsection{Dynamical quarks}\label{flavorsubsec}

One rather generic maneuver that falls into the second category just discussed (and in fact, also the first) is the one employed to add to our theories matter in the fundamental representation of the gauge group. Already in the large-$N_c$ studies of the 70s hinting at a connection \cite{thooftlargen} between non-Abelian theories and string theory, it was clear that gluons would be associated with closed strings, while quarks would be related to open strings. In the previous section we mentioned the latter identification when we introduced infinitely-massive quarks as probes of the theory, allowing us to define Wilson loops, but now we are talking about adding finitely-massive, dynamical quarks. For this we need to have open strings with finite energy, and as always, these open strings will have no choice but to be excitations of D-branes.

For the case of MSYM, for example, we can go back to the stack of $N_c$ D3-branes that we had prior to the Maldacena limit, and add alongside it a stack of $N_f$ D7-branes that are extended along the same three spatial directions as the D3s, plus four other directions, leaving two dimensions along which the two stacks can be separated. The dimensionality and orientation of the new branes is chosen so that there is no net force between the two stacks, and in this case, cuts down by half the amount of supersymmetry that is preserved by the overall arrangement. The gauge group on the D3 and D7 stacks is $U(N_c)$ and $U(N_f)$, respectively.
Referring to a string as $p$-$q$ if it starts on a D$p$-brane and ends on a D$q$-brane, excitations of our system will consist of 3-3 strings, whose modes are $N_c\times N_c$ matrices transforming in the adjoint representation of $U(N_c)$, 3-7 (and 7-3) strings, which yield $N_c\times N_f$ matrix fields transforming in the fundamental (antifundamental) of $U(N_c)$, and 7-7 strings, which yield $N_f\times N_f$ matrices that are singlets under $U(N_c)$. Note that the minimal length of the 3-7 (7-3) strings, and therefore the masses of their various modes of excitation, depends on how far apart we place the D3 and D7 stacks.

In the ultra-low-energy limit, the massless modes of the 3-3 strings give rise to $SU(N_c)$ MSYM, the lowest modes of the 3-7 (7-3) strings give rise to $N_f$ fields in the (anti-)fundamental of $SU(N_c)$, and the 7-7 strings decouple. If $\lambda\gg 1$ and $\lambda N_f\ll N_c$, then in the gravity description the D3 stack is replaced by the AdS$_5\times$S$^5$ background, but it is still appropriate to describe the D7 stack as such, in terms of its open string excitations, neglecting its backreaction on the geometry. In perturbative language, this is the analog of including gluon loops but neglecting quark loops. In lattice QCD, it is the analog of the quenched approximation, where effects of the fermion determinant are ignored.

In the end, then, the statement \cite{kk} is that IIBST on AdS$_5\times$S$^5$, with $N_f$ D7-branes that fill up AdS$_5$ from the boundary at $z=0$ down to a lowest radial position $z=z_m$, and wrap a $z$-dependent choice of S$^2$ inside the S$^5$, is dual to the theory with $\cN=2$ supersymmetry that results from coupling MSYM to $N_f$ flavors of matter fields in the fundamental, with mass $m_q=\sqrt{\lambda}/2\pi z_m$. Because of the supersymmetry, these matter fields include both fermions and scalars (forming an $\cN=2$ hypermultiplet), and in spite of this, it is conventional to refer to them collectively as quarks. The new (D7-)branes that make possible their existence are naturally known as flavor branes, to distinguish them from the color (D3-)branes that gave rise to the gluons.

More specifically, an isolated quark or antiquark (arising from a 3-7 or 7-3 string) is dual to a string that runs from the D7-branes to the AdS geometry that used to be the D3 stack. Since strings cannot end in midair, this means that the open string must reach all the way down to
(the zero-temperature horizon at)
$z\to\infty$.
In the limit where $z_m\to 0$, the D7-branes retreat to the AdS boundary and the string must reach all the way up there, so we recover the case of an infinitely massive-quark discussed in Section~\ref{quarksubsec}.

Just as it happened in that section, when we have both a quark and an antiquark the lowest energy configuration is a U-shaped string starting and ending on the D7s. This should not to be confused with the original 7-7 strings that we had before the Maldacena limit, which had nothing to do with quarks, and completely decouple when we consider ultra-low energies.\footnote{Once again, we emphasize that after the Maldacena limit is taken, energies are no longer required to be small. See footnote~5.}
The 7-7 strings that we do see after the limit are inside AdS$_5$, which means they are inside the D3-branes. They represent mesons: bound states of a quark and and an antiquark, made possible by the intervening glue.

Importantly, in the case where we have dynamical quarks, $z_m>0$ and these U-shaped strings are not obligated to reach all the way up to $z=0$ or down to $z=\infty$, so they can be of microscopic length, giving rise to light states. For $\lambda\gg 1$ we can keep only the lowest modes among these, and describe them as fields living on the D7-branes. They include the gauge field, scalars and fermions on the D7s,\footnote{These fields arising from open strings originally live on the $7+1$ dimensions spanned by the D7s. After decomposition on the S$^2\subset\;$S$^5$ wrapped by the D7s, each such field gives rise to some number of Kaluza-Klein towers of fields on AdS.} and will be dual to meson operators in the field theory, just like fields arising from closed strings and therefore living on the entire AdS geometry are dual to glueball operators. MSYM with added $\cN=2$ fundamental matter does not confine, but it does have a discrete spectrum of deeply bound mesons \cite{kmmw,eekt}, with masses $m\propto m_q/\sqrt{\lambda}\ll m_q$.

The presence of dynamical quarks and mesons makes it possible to examine many new phenomena of theoretical and phenomenological interest. One example is the interplay between the flavor branes and the black hole (\ref{schwadsmetric}) that describes a thermal gluon plasma, which leads to the existence of a critical temperature beyond which all mesons melt \cite{mateosmelt}. A second example arises when we place a quark in the plasma, and discover that its expected Brownian motion is produced on the gravity side \cite{brownian,sonteaney} by none other than the Hawking radiation emerging from the black hole along the string dual to the quark! Similar surprises are found even at zero temperature \cite{dampingtemp,holographiclessons}. Notably, the classical string on AdS knows about three important quantum properties of its dual, finitely-massive quark: it is \emph{not} pointlike (it has a finite Compton wavelength) \cite{hovdebo,dampingtemp,tmunu}; when accelerated, it is subject to a radiation-damping force expressed through a non-pathological generalization of the Lorentz-Dirac equation \cite{lorentzdirac,tail}; and when uniformly accelerated, it feels itself immersed in a thermal medium, in accord with the Unruh effect \cite{unruh}.

One can move beyond the quenched approximation by taking into account the backreaction \cite{kv} of the D7s. In this case one must face the fact that the fundamental matter makes the coupling increase logarithmically at high energies (i.e., it induces a positive beta function), meaning that the theory is no longer UV-complete, and must be studied with an explicit UV cutoff.

\subsection{Holographic QCD: the Sakai-Sugimoto Model}\label{sakaisugimotosubsec}
The top-down construction that comes closest to QCD was built in two steps, taken in papers seven years apart, by Witten \cite{wittenthermal} and by Sakai and Sugimoto \cite{ss,ss2}. Witten's setup is based on IIA string theory (IIAST), which is rather similar to IIBST, described in Section~\ref{stringsubsec}, except for the number of subindices for the $C_{M\ldots}$ gauge fields. One starts with a stack of $N_c$ D4-branes on flat spacetime, extended along spatial directions $x^1,\ldots,x^4$.
The low energy physics of this system would be described by $\cN=4$ $U(N_c)$ super-Yang-Mills in 4+1 dimensions, which as before includes massless fermions and scalars in the adjoint representation. The coupling of this five-dimensional gauge theory is determined in terms of the underlying string coupling and string length by the relation $g_5^2= 8\pi^2\gs\ls$.

One then declares the $x^4$ direction  to be a circle of radius $R_4$, $x^4\simeq x^4+2\pi R_4$.  The mass scale of the Kaluza-Klein modes associated with this circle is $M_{\mbox{\scriptsize KK}}\equiv 1/R_4$, so the theory can be viewed as approximately $3+1$ dimensional if $R_4$ is taken to be small enough ($M_{\mbox{\scriptsize KK}}$ large enough). The point of introducing the circle is using it to completely eliminate supersymmetry, by requiring that bosonic fields be periodic around it, while fermionic fields are antiperiodic.
(These are the same boundary conditions that one would use to turn on a finite temperature in the Euclidean formalism.)
At tree level this gives a mass of order $M_{\mbox{\scriptsize KK}}$ to the fermions, and scalars acquire a mass at one-loop level, so the low energy theory is pure Yang-Mills (YM) in 3+1 dimensions, with coupling $\gym^2=g_5^2/2\pi R_4$, and 't~Hooft coupling $\lambda\equiv\gym^2 N_c$.

The dual geometry is given by
\begin{equation}\label{d4metric}
ds^2=\left(\frac{u}{L}\right)^{3/2}\left[-dt^2+d\vec{x}^{\,2}+f(u)dx_4^2\right]+\left(\frac{L}{u}\right)^{3/2}\left[\frac{du^2}{f(u)}+u^2 d\Omega_4^2\right]~,
\end{equation}
where
\begin{equation}\label{f}
f(u)\equiv 1-\frac{u_0^3}{u^3}~,\qquad u_0\equiv \frac{4}{9}M_{\mbox{\scriptsize KK}}^2 L^3~,\qquad L^3\equiv\pi\gs N_c\ls^3~,
\end{equation}
and $d\Omega_4^2$ denotes the metric on the unit S$^4$. This geometry is accompanied by a non-trivial dilaton $e^{\varphi}=\gs(u/L)^{3/4}$, and $N_c$ units of flux of $F_{(4)}$ (the IIA analog of the field strength $F_{(5)}$ that we had in IIBST) through the S$^4$.
IIAST on (\ref{d4metric}) can be described by its low-energy, supergravity approximation (IIASUGRA) as long as the curvature everywhere is small in string units. Just as in MSYM, this requires $\lambda\gg 1$. But here we have a non-trivial dilaton, and we must enforce the additional requirement that the local string coupling $e^{\varphi}\ll 1$. This fails to be satisfied at very large $u$, where the appropriate description would involve supergravity in \emph{eleven} dimensions
(which is the low energy limit of M theory).

We see from (\ref{d4metric}) that the $x^4$ circle shrinks to zero size at radial position $u=u_0$, where, thanks to the specific relation between $u_0$ and $M_{\mbox{\scriptsize KK}}$ stated in (\ref{f}), the geometry caps off smoothly. Given the interpretation of $u$ as the renormalization group scale, this means that the field theory is empty in the IR region $0\le u\le u_0$, just as we expect for YM below the mass of the lightest glueball.

As in Section~\ref{stringsubsec}, we can introduce a string hanging down from the boundary on (\ref{d4metric}) to compute the quark-antiquark potential. Unlike what happened in the AdS geometry (\ref{adsmetric}), the string now can only descend down to $u=u_0$. Its tension at that depth is finite, $\sigma=\lambda M_{\mbox{\scriptsize KK}}^2/27\pi$, implying that for large quark-antiquark separations we get a linear potential, $V_{q\bar{q}}(\ell)\simeq\sigma\ell$
(in other words, the Wilson loop satisfies an area law),
signaling the linear confinement expected in YM.  To really study just YM we would want to decouple the Kaluza-Klein excitations on the $x^4$ circle by taking $M_{\mbox{\scriptsize KK}}$ very large, while holding the flux tube tension $\sigma$ fixed. But we see that this requires $\lambda\ll 1$, which is outside of the range of validity of IIASUGRA. In other words, pure YM in 3+1 dimensions is dual to the full IIAST on the stringy generalization of the background (\ref{d4metric}), over which we unfortunately have no quantitative control.

Sakai and Sugimoto \cite{ss,ss2} added fundamental matter to the Witten model by placing flavor branes on the geometry (\ref{d4metric}). The optimal choice is to use a stack of $N_f$ D8-branes filling all spatial directions except $x^4$. The lowest modes on the 4-8 and 8-4 strings then turn out to be massless fermions only (unaccompanied by scalars). Moreover, they have definite chirality: they are left-handed. On top of the D8s, Sakai-Sugimoto introduced an additional stack with $N_f$ anti-D8-branes ($\overline{\mbox{D8}}$s), which are just D8s with opposite orientation, and therefore opposite charge.
The $4$-$\bar{8}$ and $\bar{8}$-$4$ strings then give rise to massless right-handed fermions. So in this setup, the fields in the fundamental representation of the $SU(N_c)$ gauge group undoubtedly deserve to be called quarks.
As before, if $N_f\ll N_c$, then in the gravity description the D8s and $\overline{\mbox{D8}}$s can be treated as probes of the background (\ref{d4metric}), which do not generate any backreaction.

We conclude then that the (Witten-)Sakai-Sugimoto model is continuously connected to massless QCD. But unfortunately, this regime of maximal interest is literally QCD with strings attached, because it would bring into play the full IIA string theory on a highly curved, and therefore highly stringy background. At least with our present level of understanding, to be able to perform computations in the gravity description we must restrict ourselves to the large $N_c$, and more importantly, large $\lambda$ regime, where $M_{\mbox{\scriptsize KK}}\sim\Lambda_{\mbox{\scriptsize QCD}}$ and the QCD physics that we are really interested in is contaminated by the presence of the Kaluza-Klein modes on the $x^4$ circle. In other words, the field theory that gauge-gravity duality gives us quantitative access to lives in truth in $4+1$ dimensions. In spite of this limitation, experience shows that for many quantities the unwanted modes only generate small corrections, allowing many properties of the model to be in close quantitative agreement with properties of QCD \cite{ss,ss2}. A useful review can be found in \cite{rebhan}.

Particularly interesting is the geometric origin of chiral symmetry breaking. A priori, the model is invariant under $U(N_f)_L\times U(N_f)_R$ transformations.
On the gravity side these are local symmetries, the usual gauge transformations on the D8 and $\overline{\mbox{D8}}$ stacks, respectively.
In the field theory they are global symmetries, and after separating the diagonal $U(1)_B$ of baryon number and the axial $U(1)_A$, we are left with the $SU(N_f)_L\times SU(N_f)_R$   describing separate rotations of the $N_f$ species of left- and right-handed fermions. But starting with separate stacks of D8s and $\overline{\mbox{D8}}$s that hang down into the bulk from the boundary at $u\to\infty$, the lowest energy configuration will in fact be (in direct analogy with what we saw for a string and an antistring in Section~\ref{stringsubsec}) a single stack of U-shaped D8s that start and end on the boundary. And since have a single stack, we learn that the true non-Abelian symmetry of the vacuum of this field theory is just $SU(N_f)_{\mbox{\scriptsize diag}}$.
 As a result of this spontaneous breaking of chiral symmetry, Goldstone bosons appear in the spectrum of the theory: to wit, $N_f^2-1$ massless pions.
 The maximum depth to which the U-shaped D8-branes descend is
  set by the separation along $x^4$ of the D8 endpoints at $u\to\infty$.  This distance of course is not present as an independent parameter in QCD. It is customary to take it to be $\pi R_5$, so that the D8s and $\overline{\mbox{D8}}$s are at antipodal positions on the circle.

\subsection{Other top-down implementations of confinement}\label{confinesubsec}

There exist other famous top-down examples involving confinement where the field theory really lives strictly in $3+1$ dimensions. One of these is IIBST on  the Polchinski-Strassler background \cite{ps}, which is dual to the $\cN=1$ theory obtained by starting with MSYM in the UV and turning on masses for three of the fermions.
 This theory has many vacua,
 and in the gravity dual for all of them the bulk geometry terminates at some finite value of the radial coordinate $u$ or $z$ in the IR, at a location where there are explicit D5-branes or NS5-branes.
The vacuum where the $SU(N_c)$ gauge group is entirely confined, with no $U(1)$s left over, involves a single NS5-brane.

An example involving confinement and gauge symmetry breaking that is less exotic on the gravity side, but more exotic on the field theory side, is IIBST on the Klebanov-Strassler background \cite{ks}, which in the appropriate regime of parameters can be described purely in terms of IIBSUGRA. It is dual to a gauge theory that in the deep IR is precisely $SU(N_c)$ $\cN=1$ super-Yang-Mills (whose content is only gluons and gluinos), but for higher energies has gauge group $SU(kN_c)\times SU([k-1]N_c)$ and matter in the bifundamental of this group, with $k=2,3,\ldots$ growing without bound in the UV, through an infinite sequence of Seiberg dualities \cite{strasslercascade}. This behavior, known as a `duality cascade', is rather interesting from the theoretical perspective, because it in fact provides the only known example of a field theory that is well defined at arbitrarily high energies and yet does not originate from a unique CFT in the UV. It instead involves a flow that spirals past an infinite number of quasi-fixed points \cite{strasslercascade}. Also interesting is the fact that this purely four-dimensional theory can be continuously deformed \cite{bgmpz} to another famous example of a dual for $\cN=1$ SYM, the Maldacena-N\'u\~nez background \cite{mn}, which is dual to a field theory that in the UV is really six-dimensional.

\subsection{Bottom-up: AdS/QCD and Improved Holographic QCD}\label{adsqcdsubsec}

As we mentioned above, in the bottom-up approach to QCD, one uses the intuition gained from top-down holographic constructions to postulate a five-dimensional phenomenological model with just a few basic ingredients that are intended to mock up the desired properties of QCD. For confinement, for instance, one needs a geometry that, as in the Witten or Sakai-Sugimoto models, caps off in the IR, expressing the fact that the field theory is empty below some energy scale $\Lambda_{\mbox{\scriptsize IR}}$, and giving rise to a linear quark-antiquark potential. The crudest way of achieving this is to take the AdS geometry (\ref{adsmetric}) that is known to be dual to the vacuum of a CFT and by fiat truncate its $z\ge z_{\mbox{\scriptsize IR}}\equiv 1/\Lambda_{\mbox{\scriptsize IR}}$ region. This is known for obvious reasons as the hard-wall model \cite{pshard,ekss,drp}. It is instructive for initial explorations of some questions, including hard scattering \cite{pshard}, the meson spectrum and chiral symmetry breaking \cite{ekss,drp}, or a simple estimate of the deconfinement temperature \cite{herzog}, sometimes producing surprisingly good agreement with experimental data.
But for most purposes, it is too crude. For instance, its meson spectrum fails to obey the Regge behavior ($m^2$ linear in spin or radial excitation number) expected from a theory with linear confinement, and seen in actual experimental data.

A refinement is the soft-wall model \cite{kkss}, where a dilaton field is introduced that grows quadratically with the (inverted) AdS radial coordinate, $\varphi\propto z^2$. Assuming that mesons are described by an effective action with an overall factor of $e^{-\varphi}$ in front, as is the case when they are obtained from flavor D-branes in top-down models, the growing dilaton has the effect of smoothly shutting off the IR region, without having to explicitly remove it. The quadratic growth is chosen specifically to match the expected Regge behavior.
The hard- and soft-wall models, as well as their generalizations, are often referred to as AdS/QCD. (Sometimes the same denomination is applied to the entire bottom-up approach, or even to top-down models such as Sakai-Sugimoto.) One problem that they share is that the chosen backgrounds are not obtained by solving a set of bulk equations of motion that originates from an action, so one cannot carry out reliable calculations for the properties of the gluonic sector of the putative field theory.

To date, the most highly developed bottom-up model is the one formulated by Gursoy, Kiritsis and Nitti \cite{gk,gkn}, and further explored by these same authors with Mazzanti and other collaborators \cite{gkmn,ihqcdreview}. (A similar approach had been proposed by Gubser and Nellore~\cite{gn}.) It goes by the name of Improved Holographic QCD (IHQCD), and its action is that of five-dimensional Einstein gravity coupled to a dilaton (a massless scalar field),
\begin{equation}\label{ihqcdaction}
S=-M_{\mbox{\scriptsize P}}^3 N_c^2 \int d^5 x \sqrt{-g}\left[R-\frac{4}{3}(\p\varphi)^2+V(\varphi)\right]~,
\end{equation}
where $N_c$ is the number of colors and $M_{\mbox{\scriptsize P}}$ the Planck scale with its $N_c$-dependence removed (i.e., Newton's constant is really $G_N=1/16\pi M_{\mbox{\scriptsize P}}^3 N_c^2$), for convenience in the large $N_c$ limit. The motivation for the model is best explained in \cite{kiritsis}. $\lambda=e^{\varphi}$ is interpreted as (proportional to) the 't~Hooft coupling, and the key feature of the model is that the dilaton potential is chosen by hand to reproduce the expected QCD asymptotics. Namely, $V(\lambda)=(12/l^2_{\mbox{\scriptsize P}})[1+V_1\lambda+V_2\lambda^2+\ldots]$ for $\lambda\to 0$, with the $V_n$ determined by the perturbative beta function coefficients $b_n$, and $V(\lambda)\sim\lambda^{4/3}\sqrt{\ln\lambda}+\ldots$ for $\lambda\to\infty$, to obtain linear confinement in the IR. In the end, with only two phenomenological parameters, the model can successfully match many results of lattice QCD, at zero or finite temperature \cite{gkmn}. Its strength as compared with the lattice is that it allows computation of many dynamical quantities. A good review is \cite{ihqcdreview}.  Fundamental matter can be added similarly to top-down models, by introducing $N_f$ pairs of D4-branes and $\overline{\mbox{D4}}$-branes.

\section{Final comments}

In this review we have gotten a glimpse of holographic duality, and the miraculous way in which it equates ordinary field theories to theories of quantum gravity. There exists a large catalog of pairs of theories that are equated in this manner, and we have emphasized here the main examples that capture strongly-coupled QCD-like physics in a setup amenable to calculations.

It should be clear from our discussion why in all of these examples we gain access not to  QCD proper, but to gauge theories that are cousins of QCD or contain QCD coupled to additional degrees of freedom. The point is that the spectrum of QCD \emph{does not} have a large gap separating the masses of the spin $\le 2$ modes from all the rest, so we do not expect it to be dual to a theory involving just the graviton and a few other light fields (in a two-derivative action).
 This is normally expressed by saying that QCD does not have a gravity dual, but this phrasing can be misleading, because it refers specifically to a theory involving gravity that is under quantitative control, i.e., a weakly-coupled and low-curvature gravity dual. This does not preclude QCD from having a \emph{string} dual. All available evidence suggests that QCD really is completely equivalent to a string theory. We outlined one path to such a stringy description in Section \ref{sakaisugimotosubsec}.

The absence of a large gap in the QCD spectrum is directly related to the property of asymptotic freedom: it is precisely the weakly-coupled UV region that we cannot reliably describe without resorting to the full string theory framework. Just as we saw in Section \ref{sakaisugimotosubsec} for the Sakai-Sugimoto (SS) model, it happens generically that when we insist that the QCD scale $\Lambda_{\mbox{\scriptsize QCD}}$ (or equivalently, the tension $\sigma$ of the QCD string) be much smaller than the mass of the non-QCD modes that we wish to decouple (e.g., the Kaluza-Klein scale $M_{\mbox{\scriptsize KK}}$ beyond which the SS field theory really lives in $4+1$ instead of $3+1$ dimensions), we are forced to have $\lambda\ll 1$, outside of the range of validity of supergravity. Of course, the $\lambda\ll 1$ region is precisely what we can access directly with the standard perturbative expansion in the field theory, so one can attempt a hybrid approach, where the UV and IR are respectively described by the field and gravity languages  \cite{evans1,evans2,casalderreyhybrid,mateos}.

As we have seen throughout this review, in spite of the various practical limitations, many properties of the strong interaction have been modeled in terms of a gravity dual, with some degree of success. In some cases, progress has been made by systematically constructing (top-down or bottom-up) models that are closer relatives of QCD, at least for the aspects under consideration. In other cases, it has been possible to identify specific quantities whose value turns out to be the same in all (or in some large subset) of the field theories available in the holographic catalog. It then becomes plausible that the same value should be expected in QCD. Of course, in either case, extrapolation to QCD starting from a different theory is not a controlled approximation. We had emphasized this point already in the Introduction: the theories to which the gauge/gravity correspondence grants us computational access are toy models of the real-world systems of interest, and we value these toys greatly because they provide us with physical intuition far away from the familiar perturbative regime.

Beyond its usefulness as a tool to model some aspects of the strong interaction or of the vast range of materials described by strongly-coupled field theories, the gauge/gravity correspondence has forever transformed our understanding of theoretical physics, by revealing beautiful links between seemingly disparate objects and subjects. In particular, holography has erased the boundary between field theories and string theories, which previously seemed very clearly delineated. Because of its deep conceptual significance, the correspondence is in itself a very worthy object of study, and there is much ongoing work to understand mores entries of the dictionary that implements it, and the lessons it still holds for the emergence of spacetime and gravity from the lower-dimensional and non-gravitational degrees of freedom. Many more surprises are likely to follow.


\section*{Acknowledgements}

I thank Peter Hess for the invitation to write this review. I am grateful to to all who have had a hand in my ongoing learning of the gauge-gravity correspondence, particularly Curt Callan, Igor Klebanov, Larus Thorlacius, Juan Maldacena, David Gross, Shiraz Minwalla, Mart\'in Kruczenski, Elena C\'aceres, Mariano Chernicoff, Antonio Garc\'ia, Juan Pedraza, Cesar Ag\'on, Bo Sundborg, Ulf Danielsson, David Lowe, Tomeu Fiol, \O yvind Tafjord, Konstantin Savvidy, Jos\'e Edelstein, David Mateos, Veronika Hubeny, David Berenstein, Leo Pando Zayas, Mukund Rangamani, Nadav Drukker, Sangmin Lee, Vijay Balasubramanian, Mark Van Raamsdonk, Steve Gubser, Herman Verlinde, Albion Lawrence, Sumit Das, Bartek Czech, Amos Yarom, Daniel Harlow and  Aitor Lewkowycz.
This work was partially supported by Mexico's National Council of Science and Technology (CONACyT) grant 238734 and DGAPA-UNAM grant IN107115.

\end{document}